\documentstyle[preprint,aps]{revtex}
\input psfig.tex
\begin{document}
\draft
\title{What the survivors' areas do at long times?}
\author{Boris Levitan and Eytan Domany}
\date{\today}
\maketitle
\begin{abstract}
We investigate the long time behavior of the survivors' area in the scaling
state of two dimensional soap froth. We relate this problem to the recently
studied temporal decay of the fraction of Potts spins that have never been
flipped till time $t$. The results of our topological simulations 
 are consistent with the value 
$\theta=1$ for the scaling exponent of the
survivors' areas, in agreement with a recently obtained analytical result. We
find, however, that the relaxation time needed to get into the scaling regime
depends on the degree of randomness in the topological rearrangements and
becomes very large in the deterministic limit.
\end{abstract}

Coarsening of soap froth constrained between two closely spaced parallel plates
(2d froth) has attracted the attention of physicists for nearly two decades
\cite{Rev,Long}. The main reason for the current interest is the fact that such
a froth is a simple non-equilibrium system that evolves to a {\it universal
scaling state}. Many non-equilibrium systems of varying complexity share this
fascinating characteristic of their long-time evolution. Therefore one feels
that elucidating the properties of such a scaling state and the reasons for its
widely observed occurrence may further our understanding of the dynamics of
non-equilibrium systems. To gain such insights one must find ways to model the
evolution of the froth. The chosen model has to be simple enough to allow
treatment of very large systems (needed to get good statistics) without
sacrificing any important aspect of the physics. It turns out that very
different models predict {\it static} properties of the scaling state with
acceptable accuracy, whereas the system's {\it dynamic} properties are much more
model-dependent \cite{Long}. The extent to which different models succeed in
explaining known dynamic properties is of central importance in model selection.
Once reliable models are identified, they can be used to predict new properties
of the froth.

The evolution of 2d froth is governed by the strikingly simple microscopic von
Neumann law\cite{vNeumann}, which relates the rate of area change of any bubble
to the number of its sides $l$: 
\begin{equation} 
{da_l\over dt}=k(l-6),  
\label{Neumann} 
\end{equation} 
where the constant $k$ includes the physical properties of the soap films: the
surface tension and the penetrability of the gas through the film.  One of the
immediate consequences of this equation is that bubbles with $l>6$ grow, while
those with $l<6$ shrink and finally disappear. When a bubble disappears its
neighbors undergo sudden topological rearrangements (so called T2 processes)
and change their number of sides.  Knowing the areas and the number of sides of
the neighbors of a disappearing bubble does not determine uniquely the outcome
of the T2 process; which of the possible "channels" will be realized in each
particular case is determined only by the explicit geometrical configuration of
the froth just before the bubble's disappearance\cite{Weaire}. The so called
topological approach\cite{Long} does not keep track of the explicit
configuration of the froth (such as positions of all vertices); rather,
evolution is described only in terms of the areas and the connectivity matrix.
Within such an approach one needs to supplement Eq.(1) by a phenomenological 
rule that determines the outcome of the T2 processes in terms of these
variables.  

Eq.(1) implies that the total number of bubbles in the sample decreases, so
their mean area $\bar a$ grows which gives rise to a coarsening process. 
Experiments show that the coarsening froth evolves to a scaling regime
\cite{Stavans}, where $\bar a(t)\sim t$ and the distribution of the areas and
topological classes has the scaling form: $F_l(a,t)=(1/\bar a)f_l({a/ \bar a})$.  

Past theoretical and experimental studies were devoted mainly to characterizing
the $instantaneous$ pictures of the scaling state; predicting the form of the
function $f_n({a/ \bar a})$ (for modern review see refs.\cite{Rev,Long} and
references therein) as well as the topological correlations in the
froth\cite{Rivier,Aste}. Dynamical properties of the scaling state, which are
most important for model selection, were
addressed only recently, by investigating
theoretically and experimentally the behavior of $survivors$\cite{Surv}.

Survivors were defined as follows. Consider two photos of the evolving froth,
$P_i$ and $P_f$, both taken in the scaling state, at two subsequent times $t_i$
and $t_f$, with $t_i<t_f$. Let the number of cells in these two pictures be
$N(t_i)$ and $N(t_f)$ respectively ($N(t_i)>N(t_f)$). Using the information
about all the intermediate states of the froth, we can identify on the
$earlier$ picture $P_i$ all those $N(t_f)$ cells that are present in the latter
picture $P_f$. These $N(t_f)$ cells that survived till the moment $t_f$,
constitute the sub-ensemble of "survivors" at $t_i$. The statistical properties
of this sub-ensemble differ from those of the ensemble of all the bubbles on
the same photo and depend on $t=t_f-t_i$ (survival time), characterizing,
therefore, the $dynamics$ of the scaling state.

The time dependence of the survivors' topological distribution has been
investigated recently using topological simulations and mean field calculations
\cite{Surv} under the assumption of $random$ T2 processes\cite{Bnk} and it was
shown that the distribution approaches a fixed form in the long time limit. 
Subsequently, however, we discovered that the behavior of the survivors is much
more sensitive to the details of the model than the properties of all the
bubbles. Simulating a more realistic model, one that performs T2 process in a
deterministic way\cite{modC}, we found that the distribution continues to
evolve for much $longer$ times than in the case of the random model. The length
of these simulation did not suffice, however, to determine conclusively whether
the topological distribution of the survivors approaches a fixed limit. On the
other hand, mean field theory\cite{Surv} definitely predicts the existence of a
fixed form for the survivors' topological distribution as well as the
analytical results for the $q=\infty$-Potts model\cite{Sire}, which also are
consistent with convergence of the {\it area distribution} of the survivors to
a fixed form.

The long time behavior of the survivors' area distribution is of interest also
in the context of another interesting problem that has been studied intensively
recently\cite{RZ,Derr,Sience} concerning the dynamics of the Potts model. 
Starting from a random initial state, one considers the time dependence of
$r(t)$, the fraction of spins that have never flipped from the beginning of the
evolution till time $t$. It has been found that at long times 
\begin{equation}
r(t)\sim t^{-\theta} 
\end{equation} 
Similar power laws  have been observed in reaction diffusion models\cite{rdm}
and in the breath figures' growth\cite{breath} (the fraction of the area that
has not become wet till time $t$). The $q=\infty$-Potts dynamics present a good
model for soap froth simulations\cite{Pottsbubb}. The quantity that is
analogous to $r(t)$ in the soap froth is the uncrossed area $A_{uncr}(t)$; that
is, the total area that has not been crossed by any soap boundary till time
$t$. This quantity was measured recently in experiment on soap froth\cite{expt}
and the power law has been established.

One can determine a bound of $A_{uncr}(t)$ by {\it overlaying} the two pictures
$P_i$ and $P_f$. Denote by $a_i^{(k)}$ the area of survivor bubble number $k$
in $P_i$, which evolved into bubble number $k$ in $P_f$. Denote the common area
(on the overlaid pictures) of these two bubbles by $a^{(k)}_{com}(t)$; we then
have $A_{uncr}(t)\le\sum_ka^{(k)}_{com}(t)$. Since $a_{com}^{(k)}(t)\le
a_i^{(k)}(t)$, we have
\begin{equation}
A_{uncr}(t)\le \sum_ka_i^{(k)}(t) =A_s(t),
\label{est}
\end{equation} 
where $A_s(t)$ is the total area of the survivors (to time $t$), as measured on
$P_i$. Experimental observations, obtained by overlaying the raw (unpublished)
figures such as those presented in ref.\cite{Surv} indicate that many survivors
stay within their final area $a_f(t)$; hence one expects that $A_{uncr}(t)
\approx const*A_s(t)$. Experimentally, one could measure directly both $A_s(t)$
and $A_{uncr}(t)$ independently to provide a test of the extent to which this
holds. Topological simulations do not follow the movement of the boundaries
between the bubbles; hence $A_{uncr}(t)$ is not available. On the other hand
$A_s(t)$ is easy to measure, and by studying the long time behavior of $A_s(t)$
one can at least give an upper bound for $A_{uncr}(t)$, that can be compared
with $r(t)$ for the Potts model. This connection between the survivors' area
and the first flip probability serves as additional motivation to analyze the
behavior of $A_s(t)$.  

In this work we investigate the evolution of the survivors at extremely long
times. In our simulations we use a recently introduced model\cite{MyLast}; a T2
processes is assigned either at random\cite{Bnk} or deterministically
\cite{modC}, with probabilities $\alpha$ and $1-\alpha$ respectively (i.e. at
$\alpha=1$ we get the random model\cite{Bnk} while at $\alpha=0$ - the purely
deterministic model\cite{modC}). The explicit rule used for the deterministic
case is described elsewhere\cite{modC}. Simulations of this
model were in good agreement with results obtained either from
detailed simulations\cite{Yu,Yi} or experiments\cite{Rev,Stavans,Surv} on
soap froth, while the random model does not agree with some of
these results\cite{Long}.  

Extending our previous simulations\cite{Surv,modC}, we initiate an array of
$N=1000~000$ bubbles, whose topological distribution is close to that of the
scaling state; it  evolves till the number of
bubbles is about $200$. In order to reduce fluctuations we have averaged
all our results over 12 runs.  

It should be emphasized that in order to represent a system of $N$ bubbles in a
Potts simulation one needs $mN$ spins, where the minimal number of spins needed
to define a  bubble, is at least $m\approx 10$.  Running a Metropolis type
simulation of such a system till a small fraction of the domains survive, 
one needs about
$mN$ time steps of $mN$ operations, i.e. $O(m^2N^2)$ operations for a run. For
a system as large as ours this would amount to impractically long
computational time. Our $topological$ approach uses $bubbles$ as the
elementary dynamical units. By working with a time-step of order $\bar a$, we
can complete a run in $\sim N\log N$ operations. This enables us to run large
systems for extremely long times.

Starting with $N=1000~000$ bubbles, we arrived in the scaling regime with about
$N_i=640~000$ cells. At this point we number all the bubbles and we refer to
this as our initial state.  Running our simulations further, at time $t$ we
have $N(t)$ bubbles, whose ancestors, identified in the initial state,
constitute the $N(t)$ survivors. We evaluated their average topological class
$\bar n_s(t)$ and $\bar a_s(t)$, their average area (as measured in the
$initial$ state): $\bar a_s(t)=A_s(t)/N(t)$. In the scaling state  $N(t)\sim 1/t$, so we have
\begin{equation} 
A_s(t)\propto\bar a_s(t)/t 
\end{equation} 
When presenting our data, we defined the time as $t=N_i/N$. Note, that in the
scaling regime our time is proportional to the real physical time. The time
dependences of $\bar a_s(t)$, as obtained by simulations for different values
of $\alpha$, is presented on Fig. 1. For $\alpha=1$ (pure random model) $\bar
a_s(t)$ clearly approaches a constant (that is equivalent to the value
$\theta=1$ for the exponent), while for both $\alpha=0$ and $\alpha=0.5$, $\bar
a_s(t)$ continues to increase till the end of the run, in a way consistent with
approaching a constant asymptotic value.

In order to provide comvincing support for 
this asymptotic behavior we plotted in Fig. 2 the quantity
$y(t)=-\log[1- \bar a_s(t)/a_{\infty}]$ vs. $\log t$. For each $\alpha$
a constant $a_{\infty}$ was tuned
 to obtain the best fit of $y$ to a straight line. A linear dependence
$y=c_1 + \eta\log t$ is equivalent to an algebraic form $\bar
a_s(t)/a_{\infty}=1-c t^{-\eta}$, which means
\begin{equation}
A_s=\frac{C}{t} \left( 1-c_1t^{-\eta} \right)
\end{equation}
Our data, as presented in Fig. 2, indeed yields straight lines, with very good
accuracy, for all three values of $\alpha$. The value of the power $\eta$
increases with $\alpha$. In each case the linear fit breaks down at very long
times, when the "signal" (i.e. the difference $1- \bar a_s(t)/a_{\infty}$) is 
of the same order as the "noise" due to statistical fluctuations.  

A similar picture was found for the mean topological class of the
survivors, $\bar n_s(t)$ (see Fig. 3). In particularly, we note that for the
deterministic case the topological part of the problem is still evolving, even
for very long times. Analyzing this data in a way similar to that of Fig. 2 also
indicates an asymptotic power-law approach of $\bar n_s(t)$ to a constant
value. Interestingly, $\bar n_s(t)$ and $\bar a_s(t)$ approach their
asymptotic values with different powers.

On the basis of simulations of the $q=\infty$-Potts model at $T=0$ in two
dimensions, starting from random initial conditions\cite{Derr}, the value
$\theta=0.86$ was reported. This statement seems to be in conflict with our
results as well as with the analytical result of Sire and Majumdar\cite{Sire},
that yields $\theta=1$. However, looking carefully at Fig. 1 of ref. 
\cite{Derr} one sees that the local slope varies with time, apparently
approaching $\theta=1$ at long times, in a way that resembles our Fig. 1.

In a recent experiment\cite{expt} scaling behavior of $A_{uncr}(t)$ has been
observed for evolving two dimensional soap froth, with $\theta\approx 1.1$. 
That is, $A_{uncr}(t)$ was found to decay faster than our $A_s(t)\sim 1/t$. 
Since $A_s$ is only an upper bound on $A_{uncr}$, there is no contradiction
between our result and the experiment.  The latter does seem to disagree with
the prediction of ref.\cite{Sire}, which can probably be attributed to
insufficient length of the experiment. It would be interesting to see more
extensive experimental measurements of survivors' areas and topological class
evolution  in order to compare them directly with our results. Since the power
$\eta$ depends on the details of topological rearrangements, it could be a new
test of the validity of the model.

We would like to thank C. Sire and R. Zeitak for correspondence. We are also
grateful to J. Stavans and W. Y. Tam for communicating their experimental data.
This work was supported by a grants from the Germany-Israel Science Foundation
(GIF). B.L. is grateful to the Clore Foundation for financial support.

\begin{figure}
\centerline{\psfig{file=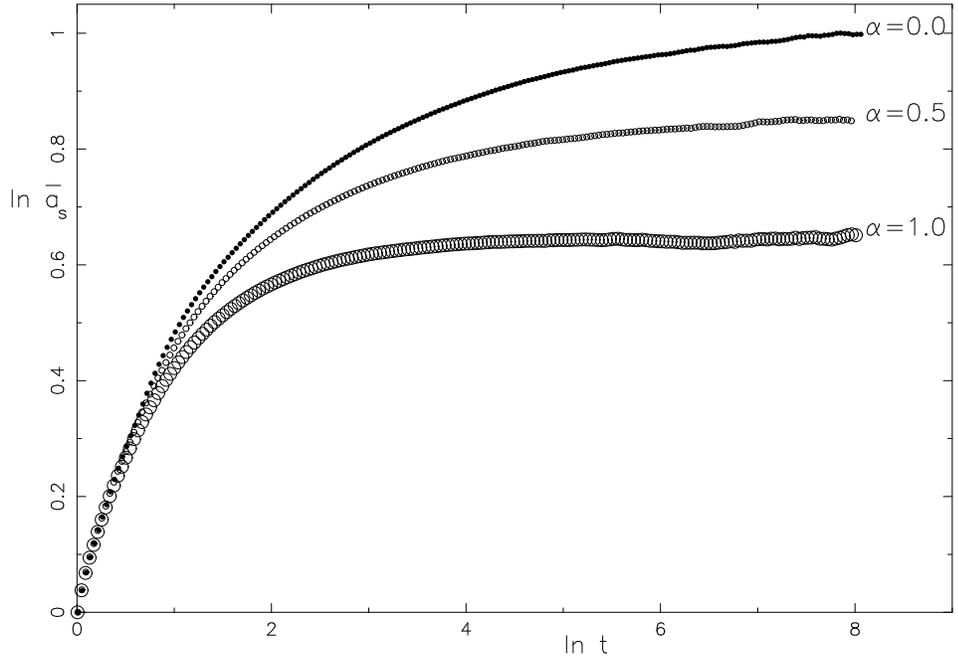,width=4.0 truein}}
\caption{
Time dependence of the mean survivors' area, $\bar a_s(t)$
as obtained by our topological simulations, for different
values of the noise parameter $\alpha$. Time is defined as $t=N_i/N$,
where $N_i$ and $N$ are the initial and the current number of bubbles in 
the system. The data have been averaged over 12 runs.
}
\end{figure}

\begin{figure}
\centerline{\psfig{file=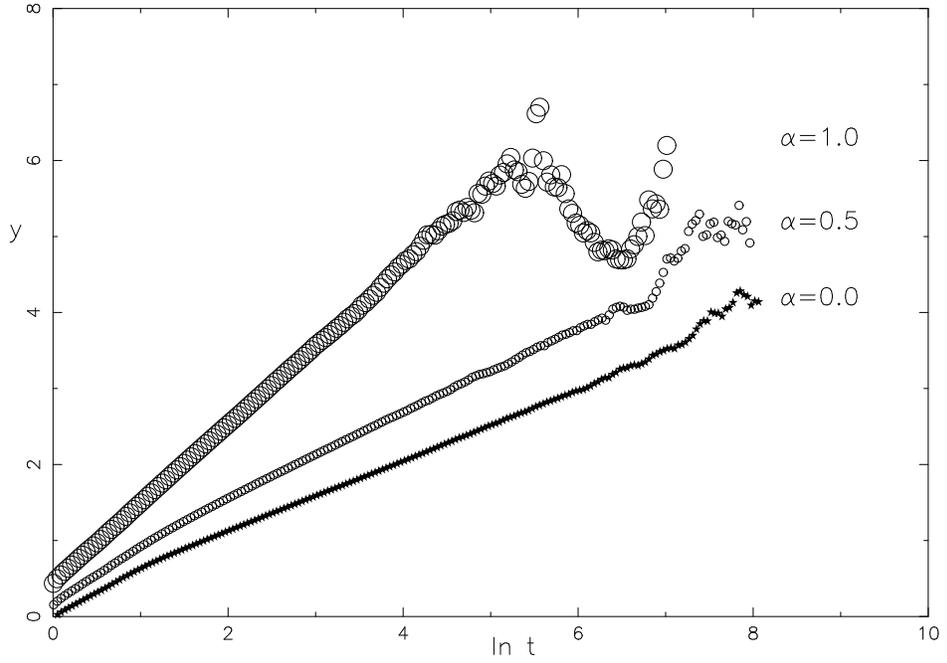,width=4.0 truein}}
\caption{
The same data as in Fig. 1: the asymptotic value $a_\infty$
was tuned so that \protect
$y(t)=-\log[1- \bar a_s(t)/a_{\infty}]$ vs. $\log t$
was approximated well by a straight line.
}
\end{figure}

\begin{figure}
\centerline{\psfig{file=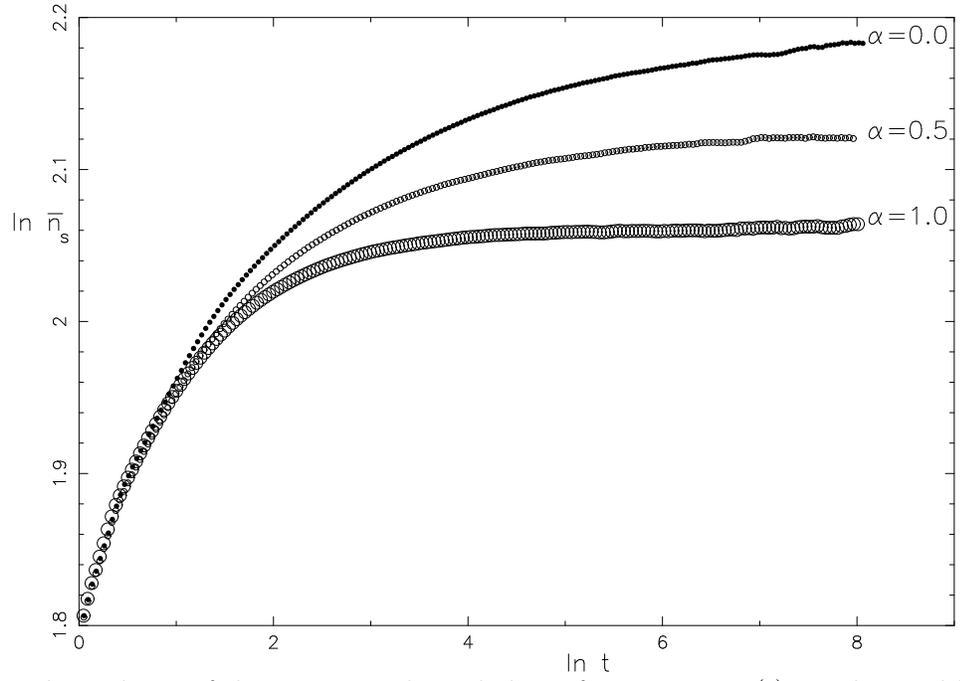,width=4.0 truein}}
\caption{
Time dependence of the mean topological class of survivors, $\bar n_s(t)$, 
as obtained by our topological simulations for different
values of the noise parameter $\alpha$. The data have been averaged over 12
runs.  The behavior of $\bar n_s(t)$ resembles that of $\bar a_s(t)$ (as
presented in Fig. 1).
}
\end{figure}


\begin{references}
\bibitem{Rev}
J. Stavans Rep. Prog. Phys. {\bf 56}, 733 (1993)
\bibitem{Long}
B. Levitan and E. Domany, Inter. Journ. of Mod. Phys. {\bf B 10}, 3765 (1996) 
\bibitem{vNeumann}
J. von Neumann, in {\it Metal interfaces}, edited by C. Herring, American 
Society of Metals, Cleveland, p. 108 (1952)
\bibitem{Weaire}
D. Weaire and J. P. Kermode, Phil. Mag. {\bf B 48}, 245 (1983);
T. Herdtle and H. Aref, J. Fluid Mech. {\bf 241}, 233 (1992)
\bibitem{Stavans}
J. A. Glazier and J. Stavans, Phys. Rev {\bf A 40}, 7398 (1989);
J. Stavans, Phys. Rev. {\bf A 42}, 5049 (1990)
\bibitem{Rivier}
N. Rivier, Phil. Mag. {\bf B 52}, 795 (1985)
\bibitem{Aste}
T. Aste, K. Y. Szeto and W. Y. Tam,
 Phys. Rev. {\bf E 54}, 5482 (1996) and references therein
\bibitem{Surv}
B. Levitan, E. Slepyan, O. Krichevsky, J. Stavans and E. Domany, 
Phys. Rev. Lett. {\bf 73}, 756 (1994)
\bibitem{Bnk}
C. W. J. Beenakker, Phys. Rev. {\bf A 37}, 1697 (1988)
\bibitem{modC}
B. Levitan and E. Domany, Phys. Rev. {\bf E 54}, 2766 (1996)
\bibitem{Sire}
C. Sire and S. Majumdar, Phys. Rev. {\bf E 52}, 244 (1995)
\bibitem{RZ}
B. Derrida, V. Hakim and V. Pasquier, Phys. Rev. Lett. {\bf 75}, 751 (1995);
B. Derrida, V. Hakim and R. Zeitak, preprint (1996)
\bibitem{Derr}
B. Derrida, P. M. C. de Oliveira and D. Stauffer,
Physica {\bf A 224}, 604 (1996) 
\bibitem{Sience}
A. Watson, Science {\bf 274}, 919 (1996)
\bibitem{rdm}
E. Ben-Naim, L. Frachebourg and P.L. Krapivsky,
Phys. Rev. {\bf E 53}, 3078 (1996)
\bibitem{breath}
S. N. Marcosmartin, D. Beysens, J. P. Bouchaud, C. Godreche and I. Yekutieli,
Physica {\bf A 214}, 396 (1996)
\bibitem{Pottsbubb}
P. S. Sahni, D. J. Srolovitz, G. S. Grest, M. P. Anderson and S. Safran,
Phys. Rev. {\bf B 28}, 2693 (1983);  
J. Wejchert, D. Weaire and J. P. Kermode,
Phil.Mag. {\bf B 53}, 15 (1986);
G. S. Grest, M. P. Anderson and D. J. Srolovitz, Phys. Rev. {\bf B 38}, 4752 
(1988)
\bibitem{expt}
W. Y. Tam, R. Zeitak, K. Y. Szeto
and J. Stavans, Phys. Rev. Lett. accepted
\bibitem{MyLast}
B. Levitan, Phys. Rev. {\bf E}, accepted
\bibitem{Yu}
H. J. Ruskin and Y. Feng, J. Phys. Condens. Matter {\bf 7}, L553 (1995)
\bibitem{Yi}
Y. Jiang, J. C. M. Mombach and J. Glazier, Phys. Rev.{\bf E 52}, 3333 (1995)

\end{references}
\end{document}